\documentclass{aa}
\usepackage{graphics}
\usepackage{graphicx}
\begin{document}

   \thesaurus{06     
              (03.11.1;  
               16.06.1;  
               19.06.1;  
               19.37.1;  
               19.53.1;  
               19.63.1)} 
   \title{The Surface Electric Field of Bare Strange Stars}


   \author{Jian Hu\inst{1}
          \and
          R.X. Xu\inst{2}
          }

   \offprints{R.X. Xu}

   \institute{Center for Astrophysics and Physics Department,
              Tsinghua University, Beijing 100084, China
    \and
    School of Physics, Peking University, Beijing 100871, China\\
    email: rxxu@bac.pku.edu.cn
    }

   \date{Received~~~~~~~~~~~~~~~~~~~~; ~~~~~accepted}

   \maketitle

\begin{abstract}

The surface electric characteristics of bare strange stars are
investigated with the inclusion of boundary effects. The
thickness of the electron layer where pairs can be spontaneously
created is calculated as a function of the bag parameters. We find
that previous estimates are representative for bag parameters
within a rather wide range, and therefore our results favor the
thermal radiation mechanism of bare strange stars advanced by V.
V. Usov.


\end{abstract}



\section{Introduction}

It is of great importance to identify strange stars; a new
window of distinguishing neutron stars and bare strange stars
(BSSs) has been proposed recently according to their sharp
differences of surface conditions (Xu, Zhang, Qiao 2001; Usov
2001a). It is therefore essential to study the surface properties
of BSSs in details, e.g., the degree of the thermal luminosity of
a hot BSS.

The surface electric field should be very strong ($\sim
10^{17}$ V/cm) near the bare quark surface of a strange star
because of the mass difference of the strange quark and the up (or
down) quark (Alcock et al. 1986), which could play an important
role in producing the thermal emission of BSSs by the Usov
mechanism (Usov 1998, Usov 2001b), because the strange quark
matter is a poor radiator of thermal photons at frequencies less
than its plasma frequency ($\sim 20$ MeV) (Alcock et al. 1986).
The basic idea of the Usov mechanism is that $e^{\pm}$ pairs are
created rapidly in a few empty quantum states with energy
$\epsilon<\epsilon_{\rm F}-2mc^2$ ($\epsilon_{\rm F}$ is the Fermi
energy, $m$ the electron mass) due to the very strong electric
field in an electron layer (with a height of $\sim 500$ fm above
quark surface); the pairs subsequently annihilate into photons
which are then thermalized
in the electron layer\footnote{ %
This layer may be optically thick for BSSs with temperature
$T\ga 10 ^9$ K.} around a BSS. This radiative mechanism has been
applied tentatively to soft $\gamma$ ray repeaters (Usov 2001c;
Usov 2001d) recently.
In addition, the strong electric field plays an essential role in
forming a possible crust around a strange star, which has been
investigated extensively by many authors (e.g., Martemyanov 1992;
Kettner et al. 1995; Huang \& Lu 1997; Phukon 2000; see Zdunik,
Haensel \& Gourgoulhon 2001  for the recent development with the
inclusion of rotating and general-relativistic effects).
Also it should be noted that this electric field has some
important implications on pulsar radio emission mechanisms (Xu \&
Qiao 1998; Xu, Qiao, Zhang 1999; Xu, Zhang, Qiao 2001).

In fact the Usov mechanism of pair production depends on many
parameters; it is therefore imperative to study the
dependence of the process on these parameters. With some typical
parameters chosen by Usov in his calculations, the resultant
thickness of the electron layer with electric field $E\ga
1.3\times 10^{16}$ V/cm (the critical field necessary for
pair production), $\Delta r_{\rm E}$, is $\sim 500$ fm. However
the proper determination of the thickness $\Delta r_{\rm E}$
should be done with the dynamical theory of the strange quark
matter.
Because of the untractable nature of the quantum chromodynamics,
some phenomenological models, i.e., the MIT bag model (e.g.,
Jensen \& Madsen 1996), the quark mass-density-dependent model
(e.g., Lugones \& Benvenuto 1995), and the quark potential model
(e.g., Dey et al. 1998), have been applied to the descriptions of
the strange quark matter .
In the bag model, $\Delta r_{\rm E}$ is a function of $\alpha_{\rm
c}$ (the coupling constant for strong interactions), $m_{\rm s}$
(the strange quark mass), and $B$ (the bag constant). Also it
should be noticed that the quark number densities, $n_{\rm i}$ (i
= u, d, s for up, down and strange quarks, respectively), are
assumed to be uniform below the quark surface, and therefore the
quark charge density $(2n_{\rm u}-n_{\rm d}-n_{\rm s})/3=V_{\rm
q}^3/(3\pi^2)$ is constant near the surface. The potential
$V_{\rm q}$ is usually chosen as 20 MeV for typical cases. However
the quark number densities should not be uniform near the surface
if boundary effects are included, since charge neutrality is
broken there.

In this paper we improve the calculation of the electric field in
the vicinity of a BSS surface, using the popular bag model and the
Thomas-Fermi model. We investigate the electric characteristics of
BSSs for different values of $\alpha_{\rm c}$. Initially we
use typical parameters for $m_{\rm s}=200$ MeV and $B^{1/4}\sim
145$ MeV. Then the thickness, $\Delta r_{\rm E}$, of the electron
layer where pairs can be created, is computed as functions of $B$,
$m_{\rm s}$, and $\alpha_{\rm c}$. It is found that Usov's
estimates are representative for bag-model parameters within a
rather wide range.
Boundary effect are also considered in this calculation.


\section{Calculation of the electric field near the quark surface}

The interesting electric properties of strange stars were first
noted by Alcock, Farhi \& Olinto (1986), who presented numerical
calculations of the electric potentials of strange stars with or
without crusts. Analytical solutions of the electron number
density and the electric field was also formulated, which could be
helpful in dealing with some physical processes near the quark
surfaces of strange stars (Xu \& Qiao 1999). In this section, the
bag model is applied to calculate the quark charge density; the
Thomas-Fermi Model is employed to find $\Delta r_{\rm E}$
dependence on $\alpha_{\rm c}$ , $m_{\rm s}$, and $B$, with the
inclusion of boundary effects.

\subsection{Equations}

In our calculation, the quarks and electrons near the surface of
the strange star keep chemical equilibrium locally; the
relation between charge density and electric potential is
described by the classical Poisson equation. The thermodynamic
potentials $\Omega_{\rm i}$ as functions of chemical potentials
$\mu_{\rm i}$ (i = u, d, s, e), $m_{\rm s}$ and $\alpha_{\rm c}$
can be found in literature (Alcock et al. 1986). In the chemical
equilibrium of weak interaction,
\begin{equation}\label{a1}
\mu_{\rm d}=\mu_{\rm s}=\mu,
\end{equation}
\begin{equation}\label{a2}
\mu_{\rm e}+\mu_{\rm u}=\mu,
\end{equation}
\begin{equation}\label{a3}
\frac{{\rm d}^2V}{{\rm d}z^2}=n_{\rm e}+\frac{1}{3}n_{\rm
d}+\frac{1}{3}n_{\rm s}-\frac{2}{3}n_{\rm u},
\end{equation}
\begin{equation}\label{a4}
n_i=-\frac{\partial\Omega_i}{\partial\mu_i},
\end{equation}
where $z$ is a measured height above the quark surface.

The quark number densities drop to zero on the surface, but they
are not uniform below the surface if the boundary effects are
considered. The chemical potential $\mu$ can be determined by the
condition that the pressure on the quark surface is zero,
$P|_{z=0}=0$. The pressure is $
P=-\sum_i\Omega_i-B
$ (Alcock et al. 1986).

The kinetic energy of electrons is equal to the electric
potential, $p_{\rm e}=V$, in the Thomas-Fermi model. The electron
number density reads (Alcock et al. 1986)
\begin{equation}\label{a5}
 n_{\rm e}=\frac{p_{\rm e}^3}{3\pi^2}=\frac{V^3}{3\pi^2}.
\end{equation}
Note that the thermodynamical potential $\Omega_i$ is defined when
the electric potential $V$ is assumed as zero. We thus replace
$\mu_i$ with $\mu_i-qV$ in Eq.(\ref{a4}), where $q$ is the
particle charges. Finally we come to,
\begin{equation}\label{a6}
\frac{{\rm d}^2V}{{\rm d}z^2}=\left\{\begin{array}{l l}
{\displaystyle \frac{V^3}{3\pi^2}+\frac{1}{3}n_{\rm
s}(V)+\frac{1}{\pi^2}\Big[\frac{1}{3}\Big(\mu+\frac{1}{3}V\Big)^3}&\\[10pt]
~~~{\displaystyle -\frac{2}{3}\Big(\mu-\frac{2}{3}V\Big)^3
\Big]\Big(1-\frac{2\alpha_{\rm c}}{\pi}\Big)},& z<0\\[10pt]
{\displaystyle \frac{V^3}{3\pi ^2},}& z\geq 0
\end{array}\right.
\end{equation}
where the complex term $n_{\rm s}(V)$ can be derived from chemical
potentials $\Omega_{\rm i}$.

The boundary conditions for Eq.(\ref{a6}) are
$$z\rightarrow-\infty:~~~V\rightarrow V_0,~~{\rm d}V/{\rm
d}z\rightarrow0;$$
$$z\rightarrow+\infty:~~~V\rightarrow 0,~~{\rm
d}V/{\rm d}z\rightarrow0.$$ where $V_0$ is the electric potential
in the deep core of the strange star, which is determined from
Eq.(\ref{a6}) by letting the left hand side equal zero.

\subsection{Results}

By integrating Eq.(\ref{a6}) in $(-\infty,V_{\rm c}]$ and $[V_{\rm
c},+\infty)$ respectively, with the boundary conditions, we get
the expression of the electric field $E=-{\rm d}V/{\rm d}z$, where
$V_{\rm c}=V|_{z=0}$. Obviously $V_{\rm c}$ is a function of
$\alpha_{\rm c}$, which can be obtained from the continuity of $E$
at $z=0$. Substituting $V_{\rm c}$ into the equations of $E$ and
then integrating them, we get the expression of $V$. For $z\geq
0$, we find
\begin{equation}
{\displaystyle V(z)=\frac{V_{\rm c}}{1+{\displaystyle \frac{V_{\rm
c}z}{\sqrt{6}\pi}}},}
\end{equation}
\begin{equation}
{\displaystyle E(z)=-\frac{{\rm d}V}{{\rm
d}z}=\frac{1}{\sqrt{6}\pi}\frac{V_{\rm c}^2}{\Big
(1+{\displaystyle \frac{V_{\rm c}z}{\sqrt{6}\pi}}\Big )^2},}
\end{equation}
\begin{equation}
{\displaystyle n_{\rm
e}(z)=\frac{V^3}{3\pi^2}=\frac{1}{3\pi^2}\frac{V_{\rm c}^3}{\Big
(1+{\displaystyle \frac{V_cz}{\sqrt{6}\pi}}\Big )^3}.}
\end{equation}

Several numerical results are obtained, which are shown in Table 1
and Figs.1-4. First, we study the electric characteristics
of a BSS for different values of $\alpha_{\rm c}$,
assuming\footnote{ Generally there is a relation between $\mu$ and
$B$, $\mu=\mu(B)$. In this assumption $B^{1/4}\sim 145$ MeV.
} %
$\mu=300$ MeV, $m_{\rm s}=200$ MeV, and the renormalization point
$\rho_{\rm R}=313$ MeV.
The quark and electron number densities \{$n_{\rm i}$\} when
$z\rightarrow -\infty$, as well as potentials of $V_0$ and $V_{\rm
c}$, are listed in Table 1 for $\alpha_{\rm
c}=0,0.1,0.3,0.5,0.7,0.9$, respectively.
\begin{table}
\begin{center}
\begin{tabular}{c|cccccc}\hline
$\alpha_{\rm c}$&0&0.1&0.3&0.5&0.7&0.9\\\hline
$B^{1/4}/$MeV&146&144&139&133&127&120\\\hline
$n_{\rm u}/10^{44}{\rm m}^{-3}$&2.86&2.71&2.41&2.10&1.80&1.48\\
\hline
$n_{\rm d}/10^{44}{\rm m}^{-3}$&3.95&3.68&3.14&2.60&2.07&1.54\\
\hline
$n_{\rm s}/10^{44}{\rm m}^{-3}$&1.77&1.74&1.68&1.61&1.52&1.42\\
\hline
$n_{\rm e}/10^{40}{\rm m}^{-3}$&13.9&11.9&7.77&4.03&1.21&0.30\\
\hline
$V_0/{\rm MeV}$&31.65&29.99&26.05&20.93&14.01&4.08\\
\hline
$V_{\rm c}/{\rm MeV}$&30.69&29.11&25.34&20.44&13.77&4.06\\
\hline
\end{tabular}
\caption{The particle number densities far below the quark
surface, and the potentials of $V_0=V(z=-\infty)$ and $V_{\rm
c}=V(z=0)$, for different values of $\alpha_{\rm c}$.}
\end{center}
\end{table}
The electric potential $V$ and field $E$ are also plotted in
Figs.1-3 for this case. We see from Fig.1 that the potential
curve becomes flatter when the coupling constant $\alpha_{\rm c}$
increases. Also it should be noted from Figs.2 and 3 that the
electric filed $E$ changes rather slowly for $z>0$, but varies
rapidly for $z<0$. This indicates the non-symmetric nature of the
electric field with respect to the quark surface of $z=0$.

 \begin{figure*}
  \includegraphics[width=8cm,height=8cm]{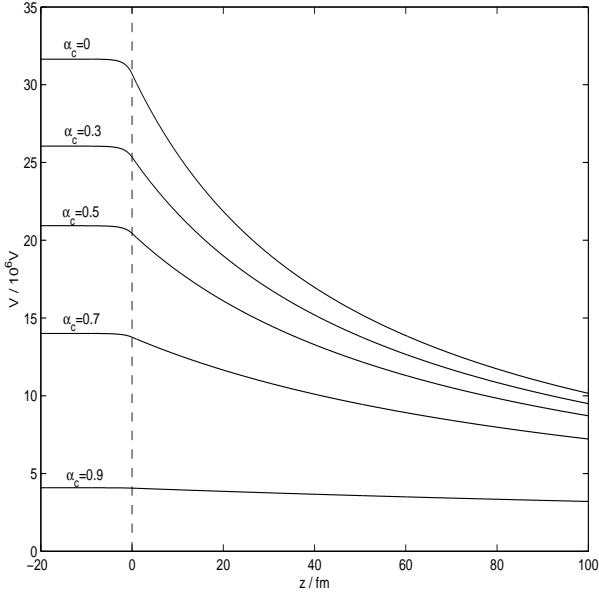}
  \hfill
  \parbox[b]{9cm}{
    \caption{Electric potential as a function of $z$ with
$\alpha_c$=0,0.3,0.5,0.7,0.9} \label{fig.1}}
  \end{figure*}

 \begin{figure*}
  \includegraphics[width=8cm,height=8cm]{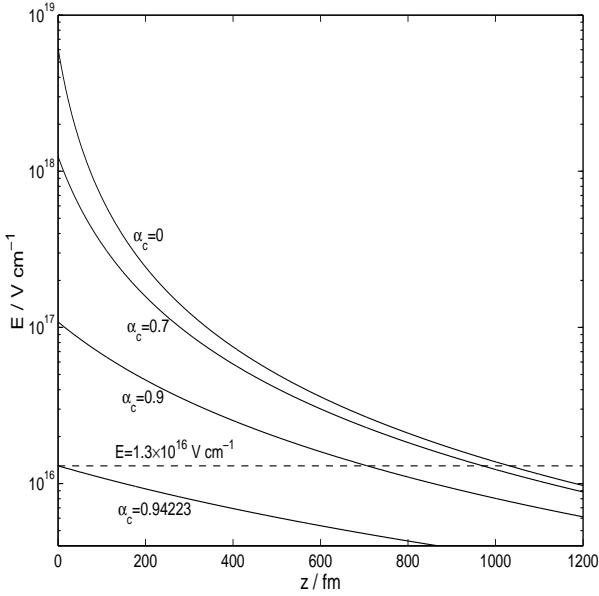}
  \hfill
  \parbox[b]{9cm}{
    \caption
    {Electric field changes slowly above the quark
surface. The coupling constant $\alpha_{\rm c}$ is chosen to be
0, 0.7, 0.9, 0.94.} \label{fig.2}}
  \end{figure*}

 \begin{figure*}
  \includegraphics[width=8cm,height=8cm]{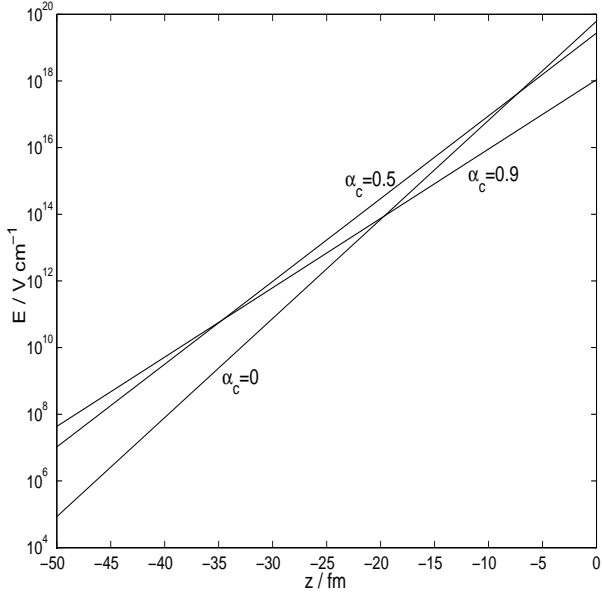}
  \hfill
  \parbox[b]{9cm}{
    \caption
    {Electric field varies sharply below the quark
surface. $\alpha_{\rm c}=0,0.5,0.9$.} \label{fig.3}}
  \end{figure*}

Secondly, we investigate numerically the thickness, $\Delta r_{\rm
E}$, of the electron layer where pairs can be created, as
functions of the bag parameters $m_{\rm s}$, $B$, and $\alpha_{\rm
c}$. The results are shown in Fig.4.
It is found that the Usov mechanism can work for bag parameters
within a rather wide range. The thickness $\Delta r_{\rm E}$ could
be large enough (i.e., $\Delta r_{\rm E}\ga 500$ fm) as long as
(1) $B$ is not too large and $m_{\rm s}$ is not too small, or (2)
$B$ is not too small and $m_{\rm s}$ is not too large. Increasing
of $\alpha_c$ has an adverse influence on the Usov
mechanism. For instance in Fig.2, the field $E$ can not exceed the
critical field $E_{\rm c}=1.3\times 10^{16}$ V/cm (necessary
for pair production) if $\alpha_{\rm c}>0.94$.
Nevertheless previous dynamical calculations show that
$\alpha_{\rm c}\ga 0.9$ is less possible for a stable strange
quark matter. In conclusion we could expect that the pair
emission process proposed by Usov might appear in nature for
acceptable bag parameters.

 \begin{figure*}
  \includegraphics[width=8cm,height=8cm]{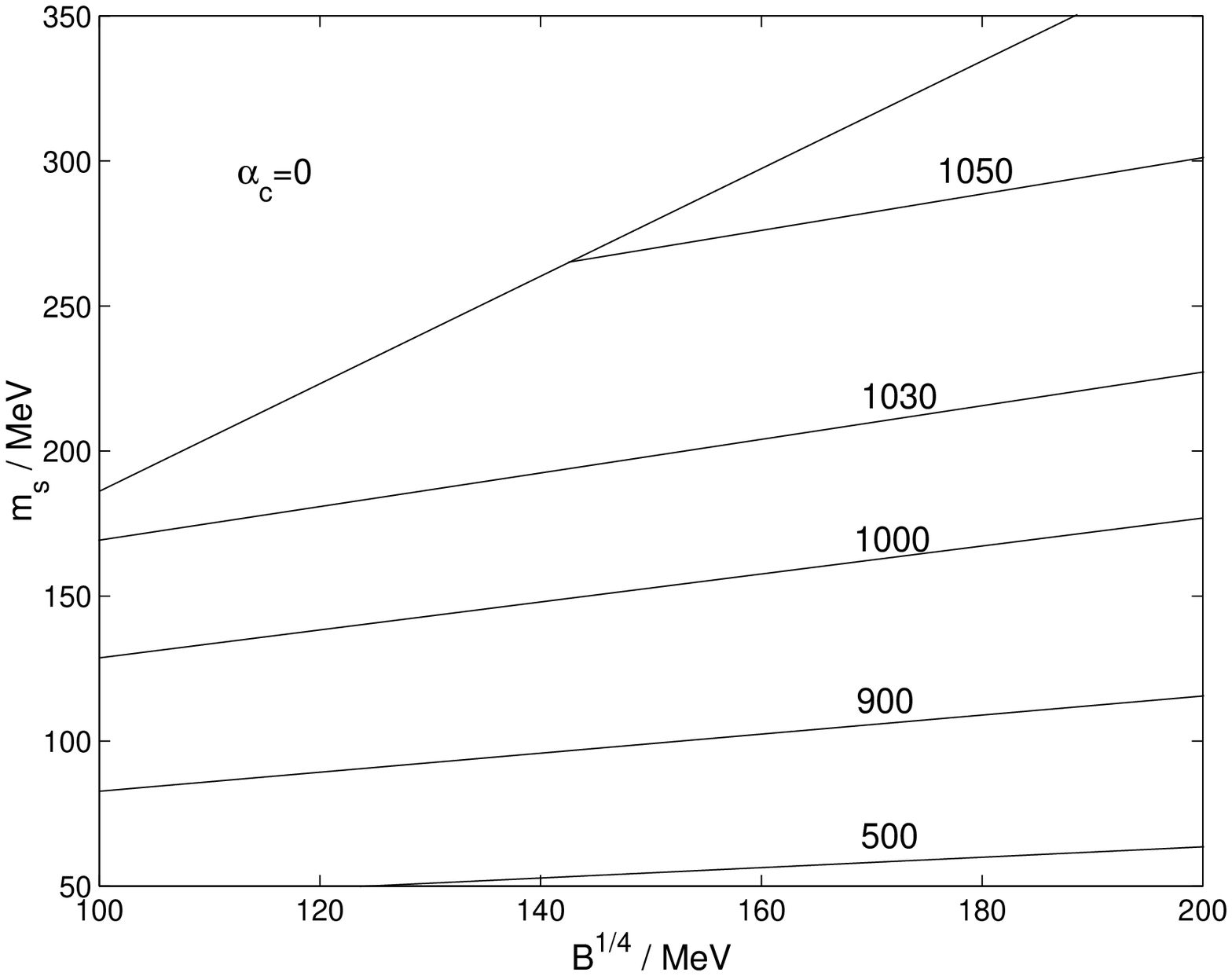}%
  \includegraphics[width=8cm,height=8cm]{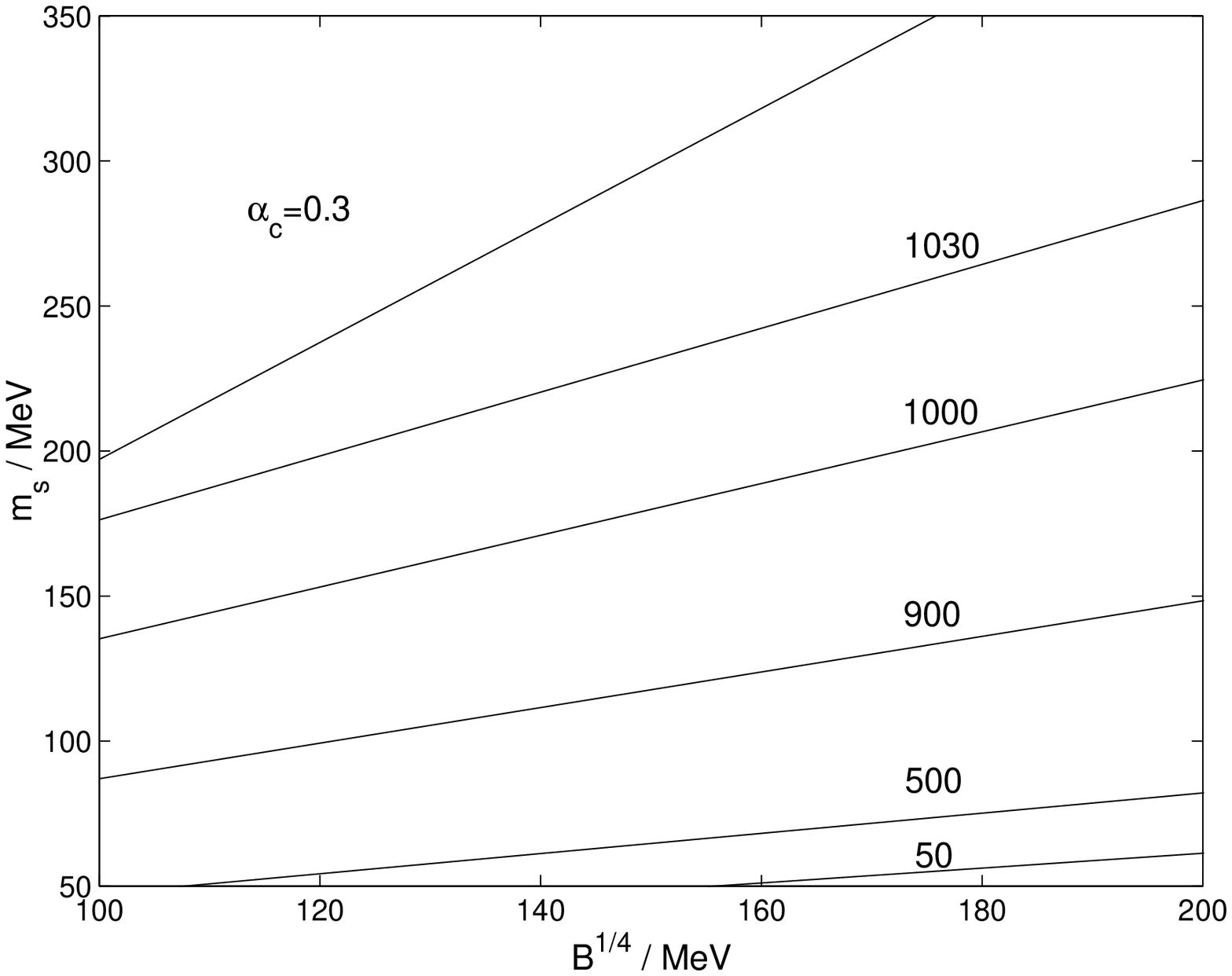}\\%
  \includegraphics[width=8cm,height=8cm]{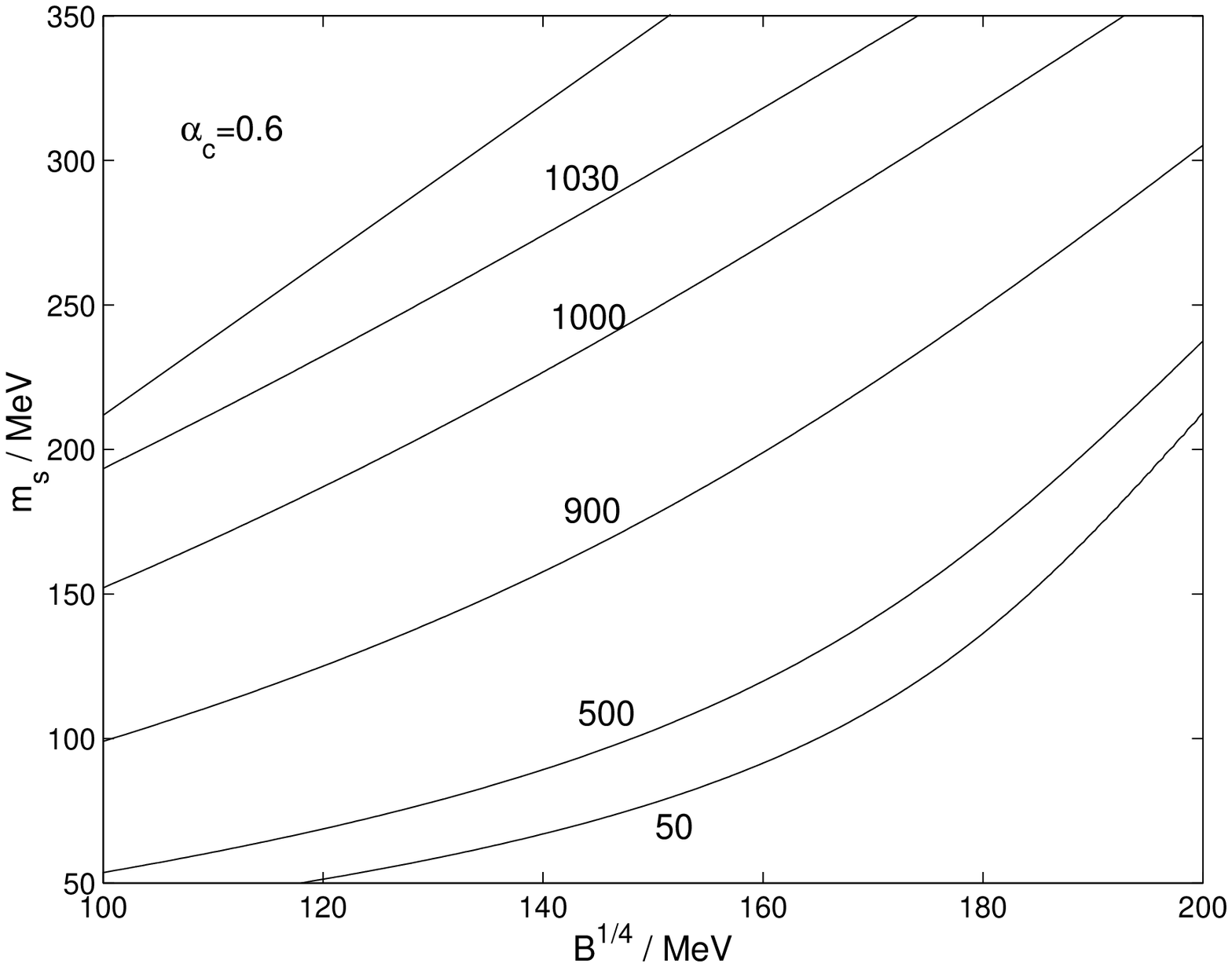}%
  \includegraphics[width=8cm,height=8cm]{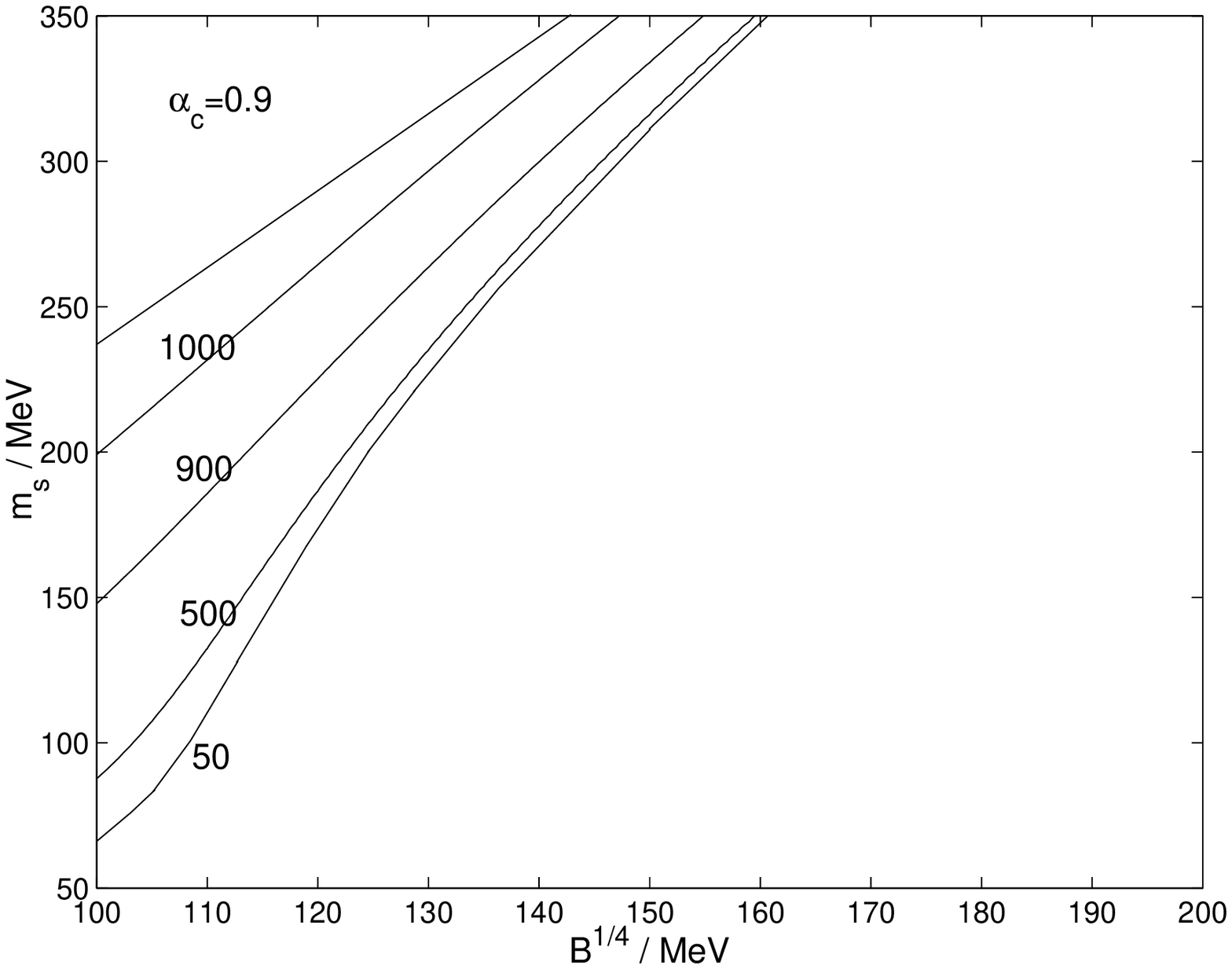}%
  \hfill
  \parbox[b]{9cm}{
    \caption{%
A set of calculated thickness $\Delta r_{\rm E}$, as functions of
bag parameters of $B$ and $m_{\rm s}$, for different coupling
constants $\alpha_{\rm c}=0,0.3,0.6,0.9$. Indicated numbers denote
the correspondence values of $\Delta r_{\rm E}$.
}%
\label{fig.4}}
  \end{figure*}
%


\section{Conclusion \& Discussion}

We have improved the calculation of the electric characteristics
of bare strange stars with the inclusion of boundary effects
(i.e., the effects of non-local neutrality near and below the
quark surface). From our calculation, we find that the Usov
mechanism can work for bag parameters within a rather wide range.

As shown in Table 1, both $V_{\rm c}$ and $V_0$, as well as their
very small difference, decrease as $\alpha_{\rm c}$ increases. Our
results on the electric potential $V(z)$ for $z<0$ is quit
different from the previous calculation given by Alcock et
al.(1986) where the boundary effects were not included (see
Fig. 1). It is shown that the strong electric field resides only
about $\sim 10$ fm below the quark surface (see Fig. 3), rather
than $\sim 10^2$ fm obtained by Alcock et al. (1986).

We can proof that $E$ is almost an exponential function of $z$
below the quark surface. Denote the right hand side of the
Eq.(\ref{a6}) as $f(V)$. As $z\rightarrow-\infty$, namely $V
\rightarrow V_0$, we have $f(V)\rightarrow0$. Approximating $f(V)$
as $f'(V_0)(V-V_0)$, we can obtain
$E=-\sqrt{f'(V_0)}\exp[\sqrt{f'(V_0)}]z$ and
$V=V_0-\exp[\sqrt{f'(V_0)}z]$ for $z < 0$.


\vspace{2mm}

\parindent=0pt

{\it Acknowledgments.}~~
This work is supported by National Nature Sciences Foundation of
China (10173002) and the Special Funds for Major State Basic
Research Projects of China (G2000077602). The authors sincerely
thank Dr. Shuangnan Zhang for his comments and the improvement of
the language.



\end{document}